\documentstyle{article}
\begin{document}
\title{
Comment on \lq\lq Giant Nernst Effect due to Fluctuating
Cooper Pairs in Superconductors" 
}

\author{
O. Narikiyo
\footnote{
Department of Physics, 
Kyushu University, 
Fukuoka 812-8581, 
Japan}
}

\date{
(Aug. 4, 2011)
}

\maketitle
\begin{abstract}
A comment is made on 
a letter (Phys. Rev. Lett. {\bf 102}, 067001) and 
a paper (Phys. Rev. B {\bf 83}, 020506) 
regarding the correct expression 
of the thermal current vertex for Cooper pairs. 
\end{abstract}
\vskip 30pt

In the Ginzburg-Landau (GL) theory~\cite{USH,LV} 
the electronic current ${\vec J}^e$ 
and the thermal current ${\vec J}^Q$ 
carried by fluctuating Cooper pairs ($T>T_c$) 
are given as 
\begin{equation}
{\vec J}^e = 2e {\vec J}, 
\end{equation}
and
\begin{equation}
{\vec J}^Q = \omega {\vec J}, 
\end{equation}
in the limit of vanishing external fields 
where 
${\vec J}$ is the particle current of Cooper pairs and 
$2e$ and $\omega$ are the charge and the energy 
carried by a Cooper pair. 
These expressions are natural and reasonable. 

\vskip 20pt

However, in a Letter~\cite{SSVG} Serbyn et al. claimed that 
the latter expression should be replaced 
by ${\vec J}^Q = 2 \omega {\vec J}$. 
Such a claim is repeated in a recent paper~\cite{LNV}. 
But there is no convincing explanation for it. 

\vskip 20pt

To give a rigorous proof of the correct expression, 
Ward identities~\cite{Nar} can play decisive role. 
The GL expressions are consistent with Ward identities. 
In the derivation of Ward identities 
the factor $2e$ or $\omega$ is the direct consequence 
of the commutation relation between the time component 
of the electronic current or the energy current 
and the Cooper-pair field. 

\vskip 30pt


\end{document}